\documentclass[prd,aps,showpacs,nofootinbib,twocolumn,superscriptaddress,
amssymb]{revtex4}

\usepackage{graphicx}
\usepackage[english]{babel}
\usepackage{amsmath}
\usepackage{amssymb}
\usepackage{amsfonts}
\usepackage{latexsym}

\newcommand{\be}{\begin{equation}}
\newcommand{\ee}{\end{equation}}
\newcommand{\bea}{\begin{eqnarray}}
\newcommand{\eea}{\end{eqnarray}}
\newcommand{\beaa}{\begin{eqnarray*}}
\newcommand{\eeaa}{\end{eqnarray*}}

\def\be{\begin{equation}}
\def\ee{\end{equation}}
\def\bea{\begin{eqnarray}}
\def\eea{\end{eqnarray}}

\begin{document}

\title{$f(T)$ gravitational baryogenesis  }

\author{V.K. Oikonomou}
\email{v.k.oikonomou1979@gmail.com}
\affiliation{Tomsk State Pedagogical University, 634061 Tomsk, Russia}
\affiliation{Laboratory for Theoretical Cosmology, Tomsk State University of Control
Systems
and Radioelectronics (TUSUR), 634050 Tomsk, Russia}

\author{Emmanuel N. Saridakis}
\email{Emmanuel\_Saridakis@baylor.edu}
\affiliation{Physics Division,
National Technical University of Athens, 15780 Zografou Campus,
Athens, Greece}
\affiliation{Instituto de F\'{\i}sica, Pontificia Universidad de Cat\'olica de
Valpara\'{\i}so,
Casilla 4950, Valpara\'{\i}so, Chile}
\affiliation{CASPER, Physics Department, Baylor University, Waco, TX 76798-7310, USA}

\begin{abstract}
We investigate how baryogenesis can occur by the presence of an $f(T)$-related
gravitational term. We study various cases of $f(T)$ gravity and we discuss in detail the
effect of the novel terms on the baryon-to-entropy ratio. Additionally, we study the
constraints imposed by the observational values of the baryon-to-entropy ratio and we
discuss how more generalized cosmologies can contribute successfully, in a viable and
consistent way, in the gravitational baryogenesis mechanism.
\end{abstract}



\maketitle

\section{Introduction}

One of the main mysteries of the standard cosmological paradigm is the explanation of the
excess of matter over antimatter, which is verified by Cosmic Microwave Background
observations  \cite{Bennett:2003bz}, as well as from Big Bang nucleosynthesis predictions
\cite{Burles:2000ju}. Gravitational baryogenesis is one of the mechanisms that have been
proposed for the generation of such baryon-anti-baryon asymmetry
\cite{Davoudiasl:2004gf,Lambiase:2006dq,Lambiase:2013haa,Lambiase:2006ft,Li:2004hh,
Pizza:2015epa,Odintsov:2016hgc}. Moreover, the gravitational baryogenesis has some
crucial effects on singular inflation (see for example \cite{oikonomou}). This mechanism
for baryon asymmetry incorporates one of Sakharov's  criteria \cite{sakharov},
and the baryon-anti-baryon asymmetry is obtained by the presence of a
$\mathcal{C}\mathcal{P}$-violating
interaction term of the form
\begin{equation}
\label{baryonassterm}
\frac{1}{M_*^2}\int \mathrm{d}^4x\sqrt{-g}(\partial_{\mu} R) J^{\mu}\, .
\end{equation}
Such a term could be acquired from  higher-order interactions in the fundamental
gravitational theory \cite{Davoudiasl:2004gf}. In particular, $M_*$ is the parameter
that denotes the cutoff scale of the underlying effective theory,
$J^{\mu}$ is the baryonic matter current, and $g$ and  $R$ are respectively the
metric determinant and the Ricci scalar. If one applies the above in the case of  a flat
Friedmann-Robertson-Walker (FRW) geometry, then  the baryon-to-entropy ratio $\eta_B/s$
is proportional to $\dot{R}$, and especially in the case where the matter
fluid corresponds to relativistic matter with equation-of-state parameter $w=1/3$ then
the net baryon asymmetry generated by the term (\ref{baryonassterm}) is zero.

In the present work we are interested in investigating the gravitational baryogenesis
mechanism in the framework of $f(T)$ gravity, which is a gravitational modification
based on the torsional (teleparallel) formulation of gravity.  In particular, in the
Teleparallel Equivalent of General Relativity (TEGR)
\cite{ein28,Hayashi79,Pereira.book,Maluf:2013gaa}
the gravitational Lagrangian is the torsion scalar $T$, and hence one can construct
torsional modified gravity by extending it to $f(T)$
\cite{Bengochea:2008gz,Linder:2010py} (see \cite{Cai:2015emx} for a review). The
interesting point is that  although  TEGR is completely equivalent with general
relativity at the equation level, $f(T)$  gravity corresponds to different
gravitational modification than $f(R)$ one, and therefore its cosmological
implications bring novel features \cite{Dent:2011zz,Geng:2011aj,Bamba:2013jqa,F-T-Inf}.

Particularly, we shall examine in detail the effects of various gravitational
baryogenesis
terms which are proportional to $\partial_{\mu}T$ or $\partial_{\mu}f(T)$. As we will
show, for the simplest choice of $f(T)$ gravity, the resulting baryon-to-entropy ratio
can
be compatible to observations, only if some parameters are chosen to be abnormally large.
Furthermore, we will constrain the functional form of more general $f(T)$ gravities which
can realize a radiation dominated Universe, and finally we shall discuss how more general
cosmologies can be contribute successfully to the gravitational baryogenesis scenario.

This paper is organized as follows: In section \ref{modela} we briefly review the
fundamental properties of $f(T)$ gravity. In section \ref{BaryogenesisF} we
discuss various gravitational baryogenesis scenarios in the
context of $f(T)$ gravity, and we examine the qualitative implications on the
baryon-to-entropy ratio which we calculate in detail for each case under study. Finally,
the conclusions follow in the end of the paper.

\section{Teleparallel and $f(T)$ gravity}
\label{modela}

In teleparallel gravity the dynamical variables are the tetrads $e^\mu_A$, which form an
orthonormal base for the tangent space at
each spacetime point. The metric is then expressed as
\begin{equation}
g_{\mu\nu}=\eta_{A B} e^A_\mu
e^B_\nu,
\label{metricvierbein}
\end{equation}
 where Greek and Latin indices  indices span the coordinate space and
tangent space respectively. Moreover, one uses the curvature-less
Weitzenb{\"{o}}ck connection  \cite{Pereira.book}
$\overset{\mathbf{w}}{\Gamma}^\lambda_{\nu\mu}\equiv e^\lambda_A\:\partial_\mu e^A_\nu$,
instead of the standard torsion-less Levi-Civita one, and hence the gravitational field is
described not by the curvature tensor but by the torsion one, which writes as
\begin{equation}
T^\rho_{\verb| |\mu\nu} \equiv e^\rho_A
\left( \partial_\mu e^A_\nu
- \partial_\nu e^A_\mu \right).
\end{equation}
The Lagrangian of the theory is the torsion scalar $T$,  constructed by
contractions of the torsion tensor as \cite{Maluf:2013gaa}
\begin{equation}
\label{Tdefscalar}
T\equiv\frac{1}{4}
T^{\rho \mu \nu}
T_{\rho \mu \nu}
+\frac{1}{2}T^{\rho \mu \nu }T_{\nu \mu\rho}
-T_{\rho \mu}{}^{\rho }T^{\nu\mu}{}_{\nu}.
\end{equation}
Similarly to $f(R)$ gravity where one extends the Einstein-Hilbert Lagrangian, namely
the Ricci scalar $R$, to an arbitrary function $f(R)$, one can generalize $T$ to
$T+f(T)$  \cite{Bengochea:2008gz,Linder:2010py,Cai:2015emx} obtaining $f(T)$ gravity as:
\begin{equation}
  S= \int d^4x e \left[
\frac{T+f(T)}{2{\kappa}^2}
\right],
\label{actiontotfT}
\end{equation}
 where
$e= \det \left(e^A_\mu \right)=\sqrt{-g}$ and $\kappa^2 =8\pi G=M_p^{-1}$ is the
gravitational constant, with $M_p$ the Planck mass.

The field equations   for $f(T)$ gravity arise by variation of the total action
$S+ S_{\mathrm{M}}$, where $ S_{\mathrm{M}}$ is the matter action, in terms of the
tetrads, namely
\begin{eqnarray}
\label{eomsgeneral}
&&\!\!\!\!\!\!\!\!\!\!\!\!\!\!\!
e^{-1}\partial_{\mu}(ee_A^{\rho}S_{\rho}{}^{\mu\nu})
(1+f_T)
-(1+f_T)  e_{A}^{\lambda}T^{\rho}{}_{\mu\lambda}S_{\rho}{}^{\nu\mu}
  \nonumber\\
&&\!\!\!\!\!\!\!\!\!\!\!\!\!\!\!
 +
e_A^{\rho}S_{\rho}{}^{\mu\nu}\partial_{\mu}({T})f_{TT}
+\frac{1}{4} e_ { A
} ^ {\nu}\left[T+f(T)\right]\! =\! \frac{{\kappa}^2}{2}
e_{A}^{\rho}\,{T^{(\mathrm{M})}}_\rho^{\verb| |\nu},
\end{eqnarray}
with $f_T=df(T)/dT$ and $f_{TT}=d^2f(T)/dT^2$, and
where the ``super-potential'' tensor $S_\rho^{\verb| |\mu\nu} = \frac{1}{2}
\left(K^{\mu\nu}_{\verb|  |\rho}+\delta^\mu_\rho
  T^{\alpha \nu}_{\verb|  |\alpha}-\delta^\nu_\rho
 T^{\alpha \mu}_{\verb|  |\alpha}\right)$ is defined in terms of the co-torsion tensor
$K^{\mu\nu}_{\verb| |\rho}=-\frac{1}{2}\left(T^{\mu\nu}_{\verb|  |\rho} - T^{\nu
\mu}_{\verb|  |\rho} - T_\rho^{\verb| |\mu\nu}\right)$. Additionally, ${
T^{({\mathrm{M}})}}_\rho^{\verb| |\nu}$ denotes the energy-momentum tensor
corresponding to $S_{\mathrm{M}}$. Note that when $f(T)=T$ one obtains
the
teleparallel equivalent of general relativity, in which case equations
(\ref{eomsgeneral}) coincide with the field equations of the latter.

Let us now apply $f(T)$ gravity in a cosmological framework. We consider a flat
 FRW background geometry with metric
\begin{equation}
ds^2= dt^2-a^2(t)\,\delta_{ij} dx^i dx^j,
\end{equation}
with $a(t)$ the scale factor, which arises from the diagonal vierbein
\begin{equation}
\label{weproudlyuse}
e_{\mu}^A={\rm
diag}(1,a,a,a).
\end{equation}
Inserting the vierbein choice (\ref{weproudlyuse}) into the field equations
(\ref{eomsgeneral}) we obtain the modified Friedmann equations
\begin{eqnarray}
\label{background1}
&&H^2= \frac{8\pi G}{3}\rho
-\frac{f}{6}+\frac{Tf_T}{3}
\\
\label{background2}
&&\dot{H}=-\frac{4\pi G(\rho+P)}{1+f_{T}+2Tf_{TT}},
\end{eqnarray}
with
$H\equiv\dot{a}/a$ is the Hubble parameter, and where a  ``dot'' denotes the
derivative with respect to  $t$. In the above equations $\rho$ and $P$
correspond to the effective energy density and the pressure of the matter content of the
Universe.
Finally, note that we have
 used   the relation
\begin{eqnarray}
\label{TH2}
T=-6H^2,
\end{eqnarray}
which according to   (\ref{Tdefscalar}) holds for an FRW Universe.

\section{Baryogenesis in $f(T)$ gravity}
\label{BaryogenesisF}

Let us now work in the torsional formulation of gravity, and consider a
$\mathcal{C}\mathcal{P}$-violating
interaction term of the form
\begin{equation}
\label{baryonasstermtorsion}
\frac{1}{M_*^2}\int \mathrm{d}^4x\sqrt{-g}\left(\partial_{\mu} (-T)\right) J^{\mu}\, .
\end{equation}
In the analysis to follow we assume that thermal equilibrium exists, thus in all
cases which we study we assume that the Universe evolves slowly from an equilibrium
state to an equilibrium state, with the energy density being related to the temperature
${\cal{T}}$ of each state as
\begin{equation}
\label{equilibrium}
\rho=\frac{\pi^2}{30}g_*\, {\cal{T}}^4\, ,
\end{equation}
where $g_*$ denotes the number of the degrees of freedom of the effectively massless
particles \cite{Davoudiasl:2004gf}. Hence,
for the $\mathcal{C}\mathcal{P}$ violating interaction of Eq.
(\ref{baryonasstermtorsion}), the induced chemical potential straightforwardly
reads  \cite{Davoudiasl:2004gf} $\mu\sim\pm\dot{T}/M_*^2$, and thus
the corresponding baryon-to-entropy ratio becomes
\begin{equation}
\label{baryontoentropyrationori}
\frac{n_B}{s}\simeq \frac{15g_b}{4\pi^2g_*}\frac{\dot{T}}{M_*^2
{\cal{T}}}\Big{|}_{\cal{T}_D}\, ,
\end{equation}
where $\cal{T}_D$ is the temperature in which the baryon current violation decouples.
Now depending on the specific torsional gravity that controls the evolution, certain
differences may occur, which we discuss in the following two subsections.

\subsection{Simple teleparallel gravity}

In simple TEGR, if the Universe is filled with a perfect fluid with constant equation of
state parameter $w=P/\rho$, the torsion scalar  is given by (\ref{TH2}), which using
(\ref{background1}) becomes
\begin{equation}
\label{torsionperffluid}
T=-16\pi G \rho\, .
\end{equation}
Interestingly enough, according to (\ref{torsionperffluid}) it can be seen that the
resulting baryon-to-entropy ratio is not zero, independently of the equation-of-state
parameter of the Universe. This is a radical contrast with general relativity, where
in the case of a radiation dominated era the
resulting baryon-to-entropy ratio is zero \cite{Odintsov:2016hgc}, since the
Einstein-Hilbert
equations are of the form
\begin{equation}\label{ricci}
R=-8\pi G (1-3w)\rho\, .
\end{equation}

Let us see what the simple form of equation (\ref{torsionperffluid}) implies for the
baryon to
entropy ratio. Firstly, by assuming a radiation dominated Universe, the energy density is
of the
form $\rho=\rho_0a^{-4}$, with $a(t)$ the scale factor. Consequently, the
differential
equation (\ref{torsionperffluid}) can be analytically solved to yield the scale
factor
\begin{equation}\label{scalesimpletors}
a(t)=\left(8\sqrt{\pi G\rho_0}t+1\right)\, ,
\end{equation}
where we have assumed that $a(0)=1$. We mention that mathematically there is an
additional solution to Eq. (\ref{torsionperffluid})
for a radiation dominated Universe, but since it is unphysical we omit it. By using
(\ref{torsionperffluid}), (\ref{scalesimpletors}) and (\ref{equilibrium}), we can
obtain the decoupling
time $t_D$ as a function of the decoupling temperature $\mathcal{T}_D$, namely
\begin{equation}\label{decouptimetemp}
t_D=\frac{3}{4\pi G}\sqrt{\frac{5}{\rho_0
g_*}}\frac{1}{\mathcal{T}_D^2}-1\, .
\end{equation}
Note that the relation (\ref{decouptimetemp}) is an exact relation and not a leading
order approximation, and this situation is unique in the cases we shall consider in this
work, since in more complicated $f(T)$ theories leading order approximations are going
to be used. Then, inserting everything in (\ref{baryontoentropyrationori}), the
resulting
baryon-to-entropy ratio $\frac{n_B}{s}$ reads
\begin{equation}
\frac{n_B}{s}\simeq -
\frac{5760 \pi ^{5/2}  G^{3/2} \,g_b\mathcal{T}_D^5 \rho^4  }{g_* M_
*^2 \rho_0^{5/2}
\left[ \pi
\mathcal{T}_D^2   \left(8   \sqrt{\pi G \rho_0}-1\right)-6
\sqrt{\frac{5}{g_*}}\right]^3}\, .
\label{sol11}
\end{equation}
We proceed by investigating how the free parameters of the theory affect the resulting
baryon-to-entropy ratio in this simple torsional theory. As a specific example we
assume that the cutoff scale $M_*$ is equal to $M_*=10^{12}$GeV, the decoupling
temperature is  $\mathcal{T}_D=M_I=2\times 10^{16}$GeV,  with $M_I$   the upper
bound for tensor-mode fluctuations constraints on the inflationary scale, and also
that $g_b\simeq \mathcal{O}(1)$, $\rho_0\simeq 10^{-6}$GeV and $g_*\simeq 106$, which is
the total number of the effectively massless particle in the Universe
\cite{Davoudiasl:2004gf}. Then, by transforming to Planck units for simplicity, the
resulting baryon-to-entropy ratio (\ref{sol11}) is found to be $n_B/s\simeq -1.02\times
10^{-42} $, which is extremely small, compared to the observed value $n_B/s\simeq
9.2\times 10^{-11}$ and more importantly it is negative, which means that this theory
predicts an
excess of anti-matter over matter, which is unphysical. The analysis shows that the
baryon-to-entropy ratio is robust
against the changes of the initial energy density $\rho_0$. The
resulting picture reveals that in order to acquire physically consistent predictions,
the simple TEGR seems not to produce a viable
baryon-to-entropy ratio, and thus a more complicated torsional term is required.

\subsection{General $f(T)$ theories}

In this subsection we extend the above discussion in the case of generalized $f(T)$
theories. As we will see, we can use the baryon-to-entropy ratio in order to constrain
the
functional form of $f(T)$ gravity. We start by considering the power-law cosmic evolution
with
scale factor
\begin{equation}\label{scalepower}
a(t)=A\,t^{\gamma}\, ,
\end{equation}
with $A,\gamma$ being a positive constants.

Let us now consider the three viable $f(T)$ cases
according to observations \cite{Nesseris:2013jea}.

\begin{itemize}

\item
 The power-law model of Bengochea and Ferraro  \cite{Bengochea:2008gz}, namely
\begin{equation}
f(T)=B\,(-T)^n,
\label{model1}
\end{equation}
with $B$ a constant and $n>1$. Then from Eq. (\ref{background1}) we
obtain that the energy density at leading order is
\begin{equation}\label{energyleadinggeneralft}
\rho \simeq \mathcal{C}\,t^{-2n}\, ,
\end{equation}
with the parameter $\mathcal{C}$ being equal to $\mathcal{C}=B 6^{n-1} (1-18 n) \gamma ^{2
n}$.
Then, the decoupling time $t_D$ as a function of the decoupling temperature
$\mathcal{T}_D$ is found to be
\begin{equation}\label{powerlaft}
t_D=\left(
\frac{\pi^2g_*}{30\mathcal{C}}\right)^{-\frac{1}{2n}}\mathcal{T}_D^{-\frac{2}{n}}\, ,
\end{equation}
and by using the expression for the torsion scalar (\ref{TH2}), the resulting
baryon-to-entropy ratio $n_B/s$ becomes
\begin{equation}\label{powerlawft}
\frac{n_B}{s}\simeq \frac{45 g_b\left( 30 \, \mathcal{C}\right)^{-\frac{3}{2\,
n}}}{g_*M_*(\pi^2g_*)
^{-\frac{2}{3n}}\mathcal{T}_D^{-\frac{6}{n}+1}}\, .
\end{equation}
Thus, by choosing $M_*=10^{12}$GeV, $\mathcal{T}_D=M_I=2\times 10^{16}$GeV, $\gamma=0.6$,
$n=5.5$
and $B=-10^{-6}$, the baryon-to-entropy ratio becomes $n_B/s\simeq 7.53 \times
10^{-11}$, which is in very good agreement with observations.
In Fig. \ref{plot1}  we present the $n$-dependence of the baryon-to-entropy ratio in a 
specific example.
\begin{figure}[h]
\centering
\includegraphics[width=20pc]{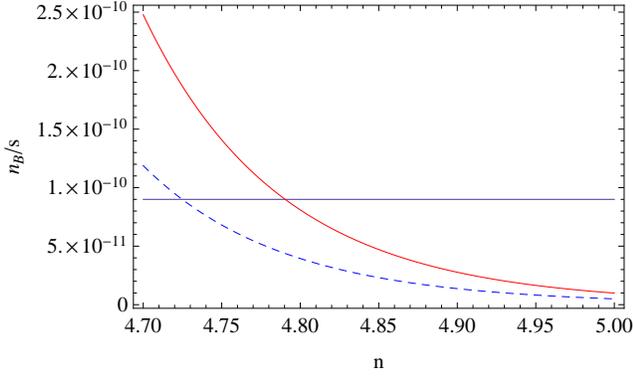}
\caption{{\it{The $n$-dependence of the baryon-to-entropy ratio $n_B/s$, for the model
$f(T)=B\,(-T)^n$, for $\mathcal{T}_D=2 \times 10^{16}$GeV, $M_*=10^{12}$GeV, $\gamma=2.5$ 
and for
$B=-10^{-6}$ (blue curve) and $B=-10^{-7}$ (red curve).}}}
\label{plot1}
\end{figure}
As it can be seen, both the parameters $B$ and $n$ affect the baryon-to-entropy ratio in
a
crucial way, and thus the baryon-to-entropy ratio may be used to constrain the
functional form of the $f(T)$ gravity.

\item

The Linder model   \cite{Linder:2010py}
\begin{eqnarray}
f(T)=B (1-e^{-p\sqrt{|T|}}),
\label{model2}
\end{eqnarray}
where $B$ and $p$ are the model parameters. In this case Eq. (\ref{background1})   at
leading order gives
\begin{equation}
\label{energyleadingKinder1}
\rho \simeq \frac{B}{6}+\frac{\gamma^2}{t^2}.
\end{equation}
The decoupling time $t_D$ as a function of the decoupling temperature
$\mathcal{T}_D$ becomes
\begin{equation}
\label{powerlaftLinder}
t_D\simeq \frac{\sqrt{30} \gamma }{\sqrt{\pi ^2 g_{*} \mathcal{T}_D^4-5 B}}\, ,
\end{equation}
and by using the expression for the torsion scalar (\ref{TH2}), the resulting
baryon-to-entropy ratio $n_B/s$ reads
\begin{equation}
\label{ftLinder1}
\frac{n_B}{s}\simeq  \frac{\sqrt{\frac{3}{10}} g_b \left(\pi ^2 g_* \mathcal{T}_D^4-5
B\right)^{3/2}
}{2 \pi ^2 \gamma  g_* M_* \mathcal{T}_D}\, .
\end{equation}
Nevertheless, for the values of the parameters $\mathcal{T}_D$ and $M_*$ we used earlier,
the resulting
baryon-to-entropy ratio is quite larger in comparison to the observed value, unless
 $\gamma$ is non-acceptably large.

\item The exponential model  \cite{Nesseris:2013jea}
\begin{eqnarray}
f(T)=B  (1-e^{-pT }),
\label{model3}
\end{eqnarray}
where $B$ and $p$ are the model parameters. In this case Eq. (\ref{background1})   at
leading order gives
\begin{equation}
\label{energyleadingKinder2}
\rho \simeq \frac{18 B \gamma ^2 p e^{\frac{6 \gamma ^2 p}{t^2}}}{t^2}\, .
\end{equation}
The decoupling time $t_D$  becomes
\begin{equation}
\label{powerlaftLinder2}
t_D=\frac{\sqrt{6} \gamma  \sqrt{p}}{\sqrt{W\left(\frac{g_* \pi ^2 \mathcal{T}_D^4}{90
B}\right)}}
\, ,
\end{equation}
where $W(z)$ is the Lambert function. By using   (\ref{TH2}) the baryon-to-entropy ratio
$n_B/s$ is
found to be
\begin{equation}
\label{ftLinder2}
\frac{n_B}{s}\simeq  \frac{5 \sqrt{\frac{3}{2}} g_b W\left(\frac{g_* \pi ^2
\mathcal{T}_D^4}{90 B}\right)^{3/2}}{2g_* M_* p^{3/2} \pi ^2 \mathcal{T}_D\gamma}\, .
\end{equation}
Hence,  by choosing $\mathcal{T}_D$ and $M_*$ as previously and also $p= 10^{-10}$, $B=1$,
and $\gamma=10^{-2}$
the resulting baryon-to-entropy ratio is $\frac{n_B}{s}\simeq 2.62 \times 10^{-11}$,
which is in very good agreement with observations. Finally, note  that
there is a wide range of values for the parameters $\gamma$ and $p$, for which we can
achieve compatibility of the baryon-to-entropy ratio with observations.

\end{itemize}

\subsection{Generalized baryogenesis term}

In this section we consider a more general baryogenesis interaction term than
(\ref{baryonasstermtorsion}), namely we extend it to
\begin{equation}
\label{baryonasstermtorsionnewterms1referee}
\frac{1}{M_*^2}\int \mathrm{d}^4x\sqrt{-g}\left[\partial_{\mu}(-T+ f(-T))\right]
J^{\mu}\,.
\end{equation}
For calculation convenience in the following we consider only the $f(-T)$ part, i.e.
 \begin{equation}
\label{baryonasstermtorsionnewterms1}
\frac{1}{M_*^2}\int \mathrm{d}^4x\sqrt{-g}\left[\partial_{\mu} f(-T)\right] J^{\mu}\, ,
\end{equation}
since the $-T$ term was considered in (\ref{baryonasstermtorsion}) and was analyzed in
the previous subsection. One can always find the results of the full $-T+ f(-T)$
consideration by adding the $\frac{n_B}{s}$ results of the separate  $-T$ and $f(-T)$
investigations.

In the case (\ref{baryonasstermtorsionnewterms1}), the resulting baryon-to-entropy ratio
becomes
\begin{equation}
\label{baryontoentropyrationoriftcase}
\frac{n_B}{s}\simeq -\frac{15g_b}{4\pi^2g_*}\frac{\dot{T}f_T(T)}{M_*^2
{\cal{T}}}\Big{|}_{{\cal T}_D}\, .
\end{equation}

Let us apply these into a power-law cosmological evolution of the form
\begin{equation}
\label{poewerlawcosmo}
a(t)=A\, t^{\gamma}\, ,
\end{equation}
with $A,\gamma>0$.  Similarly to the previous subsection, we consider the three
viable $f(T)$ cases according to observations \cite{Nesseris:2013jea}.

\begin{itemize}

\item  For the power-law model of Bengochea and Ferraro (\ref{model1}), and assuming that
  $n=2m+1$, with $m$ a positive integer, it can be shown
that the resulting baryon-to-entropy ratio at leading order is
\begin{equation}
\label{newbaryontoentropy}
\ \ \ \
\ \frac{n_B}{s}\simeq \frac{15
g_b\,2^{n+1}3^n\,B\,n\gamma^n}{4\pi^2g_*M_*}\left(\frac{\pi^2g_*}{30\mathcal{C}_1}\right)^
{\frac{2n+1}{2n}}\!\!
\mathcal{T}_D^{\frac{2(2n+1)}{n}-1},
\end{equation}
with $\mathcal{C}_1$ being equal to, $\mathcal{C}_1=2^{-1-n} 3^{-1+n} B (1-18 n)$.
Choosing the values
$M_*=10^{12}$GeV, $\mathcal{T}_D=M_I=2\times 10^{16}$GeV, $n=15$, $\gamma =1.03\times
10^{-3.8}$ and  $B=10^{-10}$, in which case we find that $n_B/s\simeq 9.24\times
10^{-11}$, which is very close to the observationally accepted
value. Alternatively, one can use $M_*=10^{12}$GeV, $\mathcal{T}_D=M_I=2\times
10^{16}$GeV, $n=3$,
$\gamma =1.23\times
10^{-23}$ and  $B=10^{-60}$, and the resulting baryon-to-entropy ratio is $n_B/s\simeq
2.014\times 10^{-11}$, but in this case, both $\gamma$ and $B$ have significantly small
values.

\item  For the Linder model (\ref{model2}), the resulting baryon-to-entropy ration is at
leading
order,
\begin{align}\label{linderftgener}
&\frac{n_B}{s}\simeq\frac{\sqrt{\frac{3}{2}} B\text{  }g_b p
\left(-5 B+g_* \pi ^2 \mathcal{T}_D^4\right) \gamma }{g_* M^* \pi ^2
\mathcal{T}_D} \\ \notag &  \times e^{-\frac{2 p \sqrt{\left(-5 B+g_* \pi ^2
\mathcal{T}_D^4\right) \gamma ^2}}{\sqrt{5}
}}\, ,
\end{align}
which is significantly small due to the presence of the exponential, regardless of the
values we
choose (unless of course we choose $p$ to be enormously small, but that is not so
appealing).

\item  For the exponential model (\ref{model3}), the baryon-to-entropy ratio at leading
order is,
\begin{equation}\label{finaleexpmod}
\frac{n_B}{s}\simeq -\frac{10 \sqrt{6} B e^{4 \gamma ^2 W\left(\frac{g_* \pi ^2
\mathcal{T}_D^4}{90
B}\right)} g_b \gamma ^2 W\left(\frac{g_* \pi ^2 \mathcal{T}_D^4}{90 B}\right)^{3/2}}{g_*
M_* \sqrt{
p} \pi ^2 \mathcal{T}_D}\, ,
\end{equation}
where $W(x)$ is again the Lambert function. Obviously this result is unphysical, since
the
resulting baryon-to-entropy ratio is negative which means that there is an excess of
anti-matter
over matter. Therefore, not all $f(T)$ gravities yield similar results.

\end{itemize}

\section{Conclusions}
\label{Conclusions}

In this paper we studied the gravitational baryogenesis scenario, generated by an $f(T)$
theory of
gravity. In the context of $f(T)$ baryogenesis, the baryon-to-entropy ratio depends on
$\dot{T}$,
and we discussed two cases of $f(T)$ theories of gravity, the case $f(T)=T$ and also
$f(T)\sim (-T)
^n$. In the first case, the resulting picture is not so appealing since in order for the
predicted
baryon-to-entropy ratio to be compatible to the observational value, some of the
parameters must
given abnormally small values. The case  $f(T)\sim (-T)^n$ is more interesting and we
investigated
which values should the parameter $n$ take in order to have compatibility with the data.
As we
showed, the variable $n$ plays a
crucial
role in the calculation of the baryon-to-entropy ratio. In both cases the interesting new
feature
is that the baryon-to-entropy ratio is non-zero for a radiation dominated Universe, in
contrast to
the Einstein-Hilbert gravitational baryogenesis scenario. Finally, we investigated how
more general
cosmologies affect the baryon to entropy ratio, and when the gravitational baryogenesis
term is of
the form $\partial_{\mu} T$, inconsistencies may occur in the theory. As we showed, the
remedy to
this issue is to modify the gravitational baryogenesis term, so that the baryon current
is
coupled
to $\partial_{\mu} f(T)$. In this way more general cosmological evolutions can be
considered and
the resulting baryon-to-entropy ratio is compatible to the observational data.

\section*{Acknowledgments}

This work is partially supported by Min. of Education and Science of Russia (V.K.O). It
is also based upon work from COST action CA15117 (CANTATA), supported by COST (European
Cooperation in Science and Technology).

\end{document}